\documentclass[prd,superscriptaddress,amsfonts,amssymb,amsmath,showpacs,twocolumn]{revtex4-2}
\usepackage{bm}
\usepackage{amsfonts}
\usepackage{latexsym}
\usepackage{graphicx}
\usepackage{amsmath}
\usepackage{palatino}
\usepackage{mathpazo}
\usepackage{textcomp}
\usepackage{float}
\usepackage{booktabs}
\usepackage{dcolumn}
\usepackage{booktabs}
\usepackage{multirow}
\usepackage{hyperref}
\hypersetup{colorlinks,citecolor=blue}
\usepackage{amsmath}
\usepackage{xcolor}
\usepackage{orcidlink}
\usepackage{epsfig}
\usepackage{caption}
\usepackage{subcaption}
\usepackage{commath}
\usepackage{natbib}
\hypersetup{colorlinks,citecolor=blue}
\hypersetup{colorlinks=true,linkcolor=magenta,filecolor=magenta,    urlcolor=blue}

\def\be{\begin{equation}}
\def\ee{\end{equation}}
\def\bea{\begin{eqnarray}}
\def\eea{\end{eqnarray}}

\begin{document}

\title{SSFDE Model: Cosmological Implications and Dynamical System Analysis}

\author{Ritika Nagpal}
\email{ritikanagpal.math@gmail.com}
\affiliation{Department of Mathematics, Vivekananda College, University of Delhi, New Delhi, India}
\affiliation{Pacif Institute of Cosmology and Selfology (PICS), Sagara, Sambalpur 768224, Odisha, India}
\author{Anil Kandel}
\email{anil.kandel3227@gmail.com}
\affiliation{Department of Physics and Astrophysics, University of Delhi, Delhi 110007, India}
\author{Ratul Mandal}
\email{ratulmandal2022@gmail.com} 
\affiliation{Department of
Mathematics, Indian Institute of Engineering Science and Technology, Shibpur, Howrah-711 103, India}
\author{Ujjal Debnath}
\email{ujjaldebnath@gmail.com} 
\affiliation{Department of
Mathematics, Indian Institute of Engineering Science and
Technology, Shibpur, Howrah-711 103, India}
\author{S. K. J. Pacif}
\email{shibesh.math@gmail.com}
\affiliation{Pacif Institute of Cosmology and Selfology (PICS), Sagara, Sambalpur 768224, Odisha, India}
\affiliation{Research Center of Astrophysics and Cosmology, Khazar University, Baku, 41 Mehseti Street, AZ1096, Azerbaijan}

\begin{abstract}
In this paper, we consider an interacting scalar field dark energy model with an exponential potential and a dark sector coupling \( Q = 3\gamma H\rho_{dm} \), which has been observationally tested using recent baryon acoustic oscillation measurements from the Dark Energy Spectroscopic Instrument  Data Release 2 , Unanchored Type Ia Supernovae, and the compressed CMB likelihood. We find that the Interacting model predicts a Hubble constant of $h = 0.659 \pm 0.0063$, deviating from the $\Lambda$CDM value by approximately $2.93\sigma$, while the Non-Interacting model shows a $3.78\sigma$ deviation. The positive coupling parameter (\( \gamma > 0 \)) further suggests a transfer of energy from dark matter to dark energy. According to the Jeffreys scale, the Interacting model shows  moderate evidence against the $\Lambda$CDM model, whereas the Non-Interacting model shows only inconclusive evidence. Further, we investigate both models through the lens of dynamical systems analysis. We formulate the cosmological evolution equations with a phenomenological interaction term and recast them into an autonomous system to study the qualitative behavior of cosmic expansion. Critical points of the system are identified and analyzed to study the corresponding cosmological dynamics. In the interacting model, we obtained five critical points, whereas in the non-interacting scenario, four distinct sets of critical points were identified. The obtained critical points, governed by cosmological parameters, represent distinct cosmic epochs, commencing from the early time stiff matter domination to late-time acceleration. Their stability is examined through linear stability analysis under appropriate physical constraints. The evolution of background cosmological parameters are also examined in terms of the dynamical system variable, and the obtained values align with observational results.
\end{abstract}

\maketitle

\section{Introduction}\label{sec_1}
In 1917, Albert Einstein introduced the cosmological constant, $\Lambda$, as a modification to General Relativity to allow for a static universe~\cite{einstein1917kosmologische}. Decades later, observations of distant Type Ia supernovae revealed that the cosmic expansion is not slowing down due to gravity, as once thought, but rather accelerating. This unexpected discovery was further supported by independent measurements of the Cosmic Microwave Background (CMB) and large scale structure (LSS), establishing that the Universe’s expansion is currently accelerating. To explain this phenomenon, a new component \textit{dark energy} (DE) was proposed. Dark energy behaves as a smooth, nearly homogeneous component with negative pressure, dominating the present-day energy budget of the Universe.

The simplest and most successful model that incorporates dark energy is the $\Lambda$Cold Dark Matter ($\Lambda$CDM) model. Within this framework, General Relativity remains valid on cosmological scales, and the total energy content of the Universe is composed of radiation, baryonic matter, cold dark matter (CDM), and a cosmological constant $\Lambda$, representing dark energy with a constant energy density. The evolution of the homogeneous and isotropic universe under $\Lambda$CDM is governed by the Friedmann equations:
\begin{equation}
\begin{aligned}
& H^2 = \left(\frac{\dot{a}}{a}\right)^2 = \frac{8\pi G\rho}{3}, \
& \frac{\ddot{a}}{a} = -\frac{4\pi G}{3}\left(\rho + 3p\right),
\end{aligned}
\end{equation}
where $H$ is the Hubble parameter, and $\rho$ and $p$ denote the total energy density and pressure, respectively. Accelerated expansion ($\ddot{a}>0$) requires an effective equation of state (EoS) $\omega = p/\rho < -1/3$. The cosmological constant corresponds to a dark energy component with $w_\Lambda = -1$.

Despite its simplicity and empirical success, $\Lambda$CDM faces several conceptual and observational challenges. Theoretically, it suffers from the cosmological constant and cosmic coincidence problems~\cite{martin2006testing,frieman2008dark,linder2007mirage}. Observationally, persistent anomalies have emerged most notably the $H_0$ and $\sigma_8$ tensions, large-angle CMB anomalies, and a possible cosmic dipole~\cite{nesseris2006evolving,wang2018evolution,heisenberg2022can,perivolaropoulos2022challenges,abdalla2022cosmology,di2020interacting,pan2019interacting,di2025cosmoverse,Capozziello:2023ewq,Capozziello:2024stm}. These suggest that the standard $\Lambda$CDM picture might be incomplete, motivating the exploration of extensions involving dynamical dark energy.

Dark energy models are commonly classified according to their EoS parameter $\omega$:
\begin{itemize}
\item $\omega = -1$: Cosmological constant ($\Lambda$).
\item $\omega > -1$: Quintessence dark energy.
\item $\omega < -1$: Phantom dark energy, often violating the null energy condition~\cite{hawking2023large,qiu2008null,moghtaderi2025much}.
\item $\omega$ crosses $-1$: Quintom dark energy, with evolving $\omega(z)$ that transitions across the cosmological constant boundary.
\end{itemize}
A widely used phenomenological parameterization for evolving dark energy is the Chevallier–Polarski–Linder (CPL) form~\cite{chevallier2001accelerating,linder2003exploring}:
\begin{equation}
\omega(z) = \omega_0 + \omega_a(1-a),
\end{equation}
where $\omega_0$ and $\omega_a$ describe the present value and evolution rate of the dark energy EoS.

Between 2003 and 2012, the Wilkinson Microwave Anisotropy Probe (WMAP) provided increasingly precise constraints on dark energy, consistently supporting $\Lambda$CDM. The first-year WMAP (WMAP1) analysis reported $w = -0.98 \pm 0.12$~\cite{spergel2003first}, followed by WMAP3 with $w = -0.967^{+0.073}_{-0.072}$~\cite{spergel2007three}. WMAP5 yielded $\omega = -1.00^{+0.12}_{-0.14}$, while allowing a time-dependent form gave $w_0 = -1.06 \pm 0.14$ and $\omega_a = 0.36 \pm 0.62$~\cite{komatsu2009five}. WMAP7~\cite{jarosik2011seven} found $\omega = -1.10 \pm 0.14$, and WMAP9~\cite{bennett2013nine} concluded with $w = -1.073^{+0.090}_{-0.089}$. All these results were consistent with $w = -1$, confirming the cosmological constant as an excellent fit to the data when combined with BAO, Type Ia supernovae, and $H_0$ measurements.

The Planck mission refined these findings with higher precision. Planck 2013~\cite{ade2014planck} obtained $w = -1.13^{+0.13}_{-0.14}$ (Planck + WMAP9 + SNLS), lying slightly within the phantom regime. The Joint Light-curve Analysis (JLA, 2014)~\cite{betoule2014improved} significantly improved supernova calibration, reducing earlier discrepancies and yielding results consistent with $\Lambda$CDM. In Planck 2015~\cite{ade2016planck}, which adopted JLA as its baseline supernova dataset, the EoS constraint became $w = -1.006^{+0.085}_{-0.091}$ in excellent agreement with a cosmological constant.

More recent datasets continued to test these conclusions with unprecedented precision. The Pantheon~\cite{scolnic2018complete}, DES Y1~\cite{troxel2018dark,abbott2018dark}, Planck2018~\cite{aghanim2020planck}, and DES Y3~\cite{abbott2022dark,abbott2023dark} analyses all found $w$ close to $-1$, though a subtle shift toward dynamical behavior began to appear. The Pantheon+ compilation~\cite{brout2022pantheon} reported $w = -0.90 \pm 0.14$ using supernovae alone and $w = -0.978^{+0.024}_{-0.031}$ when combined with CMB and BAO data. The Union3 sample~\cite{rubin2025union} strengthened this trend, showing mild ($1.7$–$2.6\sigma$) tension with $\Lambda$CDM and preferring $w_0 > -1$ and $w_a < 0$, suggestive of an evolving dark energy consistent with a Quintom-B scenario.

In 2024, the Dark Energy Survey (DES) Year 5 results~\cite{abbott2024dark} confirmed that both supernova-only and combined analyses favored $w > -1$ at about the $1\sigma$ level. Around the same time, the Dark Energy Spectroscopic Instrument (DESI) released its first-year BAO results~\cite{adame2025desi}, which deviated from $\Lambda$CDM at levels of $2.5$–$3.9\sigma$, depending on the combination of datasets. The subsequent DESI DR2~\cite{abdul2025desi} provided even stronger evidence, excluding $\Lambda$CDM at $3.1\sigma$ (DESI + CMB), and at $2.8\sigma$, $3.8\sigma$, and $4.2\sigma$ when combined with Pantheon+, Union3, and DES Y5, respectively. These findings collectively point toward a dynamical dark energy component whose EoS evolves across the phantom divide. The far-reaching implications of these findings have prompted extensive follow-up investigations into the underlying physics of dark energy \cite{choudhury2024updated,choudhury2025cosmology,tada2024quintessential,berghaus2024quantifying,park2024using,yin2024cosmic,cortes2024interpreting,lodha2025desi,carloni2025does,croker2024desi,mukherjee2024model,roy2025dynamical,wang2024dark,orchard2024probing,giare2024robust,dinda2025model,jiang2024nonparametric,rebouccas2025investigating,bhattacharya2024cosmological,pang2025constraints,ramadan2024desi,efstathiou2025evolving,cortes2025desi,wolf2025robustness,huang2025desi,tsedrik2025interacting,gao2024evidence,giare2025dynamical,chakraborty2025desi,giare2025overview,ye2025nec,lopez2025crosschecking,colgain2025much,ye2025hints,silva2025new,park2024w_0w_a,fazzari2025cosmographic,wu2025observational,chaudhary2025impact,nagpal2025late,chaudhary2025does,chaudhary2025evidence,chaudhary2025lambda,chaudhary2025probing,capozziello2025evidence,kumar2025probing,braglia2025exotic,silva2025testing,li2025exploring,mishra2025braneworld,mazumdar2025constraint,liu2025torsion,gialamas2025quintessence,luciano2025constraints,van2025compartmentalization,mukherjee2025new,moffat2025dynamical,ye2025tension,scherer2025challenging,dinda2025physical,mirpoorian2025dynamical,toomey2025theory,adam2025comparing,plaza2025probing,yang2025dark,petri2025dark,arora2025dynamical,ishak2025fall,goldstein2025monodromic,qiang2025new,li2025reconstructing,an2025topological,wang2025model}.

Such results have far-reaching implications for cosmology, motivating theoretical extensions to $\Lambda$CDM that allow for an evolving dark energy component. Among the most actively studied approaches are the $\omega$CDM and $\omega_0\omega_a$CDM parameterizations, interacting dark energy (IDE) models, scalar field scenarios, and self-interacting scalar field dark energy (SSFDE) models. In particular, IDE and SSFDE frameworks naturally accommodate a dynamical equation of state that can reproduce the evolving dark energy behavior indicated by recent DESI observations. These models provide a flexible and physically motivated framework for exploring deviations from the cosmological constant and assessing their consistency with current high-precision data.

Motivated by these recent developments and the DESI DR2 results, this study explores SSFDE models as viable alternatives to $\Lambda$CDM. Specifically, we investigate whether these models potentially including interactions between dark matter and dark energy can reconcile current cosmological tensions while remaining consistent with high precision observational data. Following the statistical analysis framework outlined in~\cite{nagpal2025}, we perform a comprehensive examination of their background evolution, stability, and parameter constraints, evaluating their ability to extend the standard cosmological paradigm.

This paper is organized as follows. Section \ref{sec_1} provides the motivation for exploring interacting and self-interacting dark energy models as natural extensions of $\Lambda$CDM, particularly in light of recent observational evidence for a dynamically evolving dark energy component. Sections \ref{sec_2} develops the theoretical framework by formulating the field equations with a phenomenological interaction term, describing different interaction scenarios, and outlining the solution methods. Section \ref{sec_3} presents the observational datasets and methodology used in this work, including the application of Markov Chain Monte Carlo (MCMC) techniques. Section \ref{sec_4} reports the dynamical systems approach to our considered work and also analyze the results together with the physical viability of the models based on statistical diagnostics. Section \ref{sec_5} summarizes the key findings and concludes the paper with final remarks\\\\

\section{Field Equations with Cosmic Matter-Energy Interactions}\label{sec_2}

In a spatially flat Friedmann--Lemaître--Robertson--Walker (FLRW) universe, the gravitational field equations (Einstein’s equations) relate the expansion rate to the total energy content. The FLRW metric can be written as $ds^2 = -dt^2 + a^2(t)d\mathbf{x}^2$ (with $a(t)$ the cosmic scale factor), and the Hubble parameter is defined as $H \equiv \dot{a}/a$. For a flat universe ($k=0$) containing radiation, baryons, cold dark matter (CDM), and dark energy (DE), the Friedmann equation is 
\begin{equation}\label{eq:friedmann}
H^2 = \frac{8\pi G}{3}\Big(\rho_r + \rho_b + \rho_c + \rho_{DE}\Big)\,,
\end{equation}
where $\rho_r$, $\rho_b$, $\rho_c$, and $\rho_{DE}$ denote the energy densities of radiation, baryonic matter, CDM, and dark energy, respectively. The acceleration equation (second Friedmann equation) is 
\begin{equation}\label{eq:accel}
\frac{\ddot{a}}{a} = -\frac{4\pi G}{3}\Big(\rho_r + \rho_b + \rho_c + \rho_{DE} + 3p_r + 3p_b + 3p_c + 3p_{DE}\Big)\,,
\end{equation}
with $p_i$ the pressure of each component.  For radiation $p_r = \rho_r/3$, for pressureless matter (baryons and CDM) $p_b = p_c = 0$, and for dark energy $p_{DE} = w_{DE}\,\rho_{DE}$ where $w_{DE}$ is the (generally time-dependent) equation-of-state parameter (e.g., $w_{DE}=-1$ for a cosmological constant). Equations (\ref{eq:friedmann}) and (\ref{eq:accel}) are the Einstein field equations specialized to the homogeneous FLRW cosmology.

In the absence of any interactions between these components (the standard $\Lambda$CDM scenario), each fluid’s energy-momentum tensor is separately conserved. The covariant conservation law $\nabla^\mu T_{\mu\nu}^{(i)}=0$ for each component $i$ leads to the continuity equation
\begin{equation}\label{eq:continuity}
\dot{\rho}_i + 3H(\rho_i + p_i) = 0~,
\end{equation}
for $i \in \{r, b, c, \text{DE}\}$ (in the non-interacting case). In particular, radiation (with $p_r=\rho_r/3$) redshifts as $\rho_r \propto a^{-4}$ and non-relativistic matter (baryons or CDM with $p=0$) as $\rho \propto a^{-3}$, consistent with Eq.~(\ref{eq:continuity}). Dark energy with a fixed $w_{DE}$ would evolve as $\rho_{DE} \propto a^{-3(1+w_{DE})}$ if it does not interact with other sectors.

We now extend this framework to scenarios in which CDM and dark energy interact by exchanging energy. Such interacting dark sector models have been proposed as a way to address the "cosmic coincidence problem” (the near-equality of $\rho_c$ and $\rho_{DE}$ today), and they have been extensively studied in the literature (see, e.g., Amendola 2000; Zimdahl et al. 2001; and Wang et al. 2016 for a review). In an interacting scenario, the total energy-momentum of the dark sector is still conserved (as required by Einstein’s equations), but the CDM and DE components are no longer conserved individually. Instead, one can write 
\begin{equation}\label{eq:Qcovariant}
\nabla^\mu T_{\mu\nu}^{(c)} = Q_{\nu}\,, \qquad 
\nabla^\mu T_{\mu\nu}^{(DE)} = -\,Q_{\nu}\,,
\end{equation}
where $Q_{\nu}$ is the interaction four-vector that quantifies the energy-momentum transfer between CDM and dark energy. Eq~ (\ref{eq:Qcovariant}) ensures that $\nabla^\mu (T_{\mu\nu}^{(c)} + T_{\mu\nu}^{(DE)}) = 0$, so that the total energy-momentum of the dark sector is conserved at all times. We assume that the interaction does not involve significant momentum exchange in the cosmic rest frame (i.e. $Q_{\nu}$ is parallel to the 4-velocity $u_{\nu}$ of the comoving frame). In that case $Q_{\nu}$ has only a time component: $Q_{\nu} = (Q, \mathbf{0})$, where $Q=Q(t)$ is the time-dependent rate of energy transfer.

Taking the time-component ($\nu=0$) of Eq.~(\ref{eq:Qcovariant}) yields the modified continuity equations for CDM and dark energy in the interacting case:
\begin{equation}\label{eq:continuity_c}
\dot{\rho}_c + 3H\,\rho_c = Q~,
\end{equation}
\begin{equation}\label{eq:continuity_DE}
\dot{\rho}_{DE} + 3H(1+w_{DE})\,\rho_{DE} = -\,Q~,
\end{equation}
consistent with $\nabla^0 T_{0}^{\,(c)} = -\,\nabla^0 T_{0}^{\,(DE)} = Q$. Equations (\ref{eq:continuity_c}) and (\ref{eq:continuity_DE}) show that the energy lost (or gained) by one component is gained (or lost) by the other, so that the total $\rho_c+\rho_{DE}$ evolves according to the standard conservation law. In our sign convention, a positive $Q$ corresponds to a transfer of energy from dark energy to dark matter (i.e. CDM gains energy while DE decays). The other components (radiation and baryons) are assumed to be uncoupled and thus obey their usual continuity equations ($Q=0$ for those fluids). Adding Eqs. (\ref{eq:continuity_c}) and (\ref{eq:continuity_DE}) indeed recovers $\dot{\rho}_c + \dot{\rho}_{DE} + 3H(\rho_c + \rho_{DE} + p_{DE})=0$, consistent with the global conservation law (Bianchi identity).

The introduction of an interaction term $Q$ in the dark sector is a phenomenological extension of the standard cosmological model, and various functional forms for $Q(t)$ have been explored in the literature \cite{Amendola2000,Zimdahl2001,Wang2005,Valiviita2008}. Common prescriptions include couplings proportional to the energy density of one or both dark components, such as $Q \propto H \rho_{DE}$, $Q \propto H \rho_c$, or $Q \propto H(\rho_c + \rho_{DE})$ \cite{Guo2007,He2009,Clemson2012}. For instance, a widely studied case assumes a linear dependence on the CDM density, $Q = \beta H \rho_c$, where $\beta$ is a dimensionless coupling parameter \cite{Amendola2000,Valiviita2008,Clemson2012}.

\begin{equation}\label{eq:Q_example}
Q = 3H\,\gamma \,\rho_{DE}\,,
\end{equation}
with $\gamma$ being a dimensionless coupling constant \cite{Amendola2000,Zimdahl2001,Valiviita2008}. This form, originally introduced in coupled quintessence models, represents a transfer of energy proportional to the DE density. A positive $\gamma$ implies that dark energy decays into CDM, while $\gamma<0$ corresponds to energy transfer in the opposite direction.

The specific form of $Q$ determines the time evolution of $\rho_c(t)$ and $\rho_{DE}(t)$ through the continuity equations (\ref{eq:continuity_c})–(\ref{eq:continuity_DE}). While the individual components are not separately conserved, the total conservation law is always maintained:
\[
\dot{\rho}_c + \dot{\rho}_{DE} + 3H(\rho_c + \rho_{DE} + p_c + p_{DE}) = 0.
\]

Several phenomenological prescriptions for $Q$ have been extensively studied. The three most common are
\[
Q = 3H \gamma \rho_{DE}, \qquad 
Q = 3H \gamma \rho_c, \qquad
Q = 3H \gamma (\rho_c + \rho_{DE}),
\]
corresponding respectively to couplings with DE only, CDM only, and the total dark sector density \cite{Amendola2000,Zimdahl2001,He2009,Valiviita2008,Guo2007,Clemson2012}. Other generalizations include forms involving $\dot{\rho}_{DE}$ or decay-rate type couplings for CDM \cite{Olivares2005,Bohemer2008}. Each of these interaction laws produces distinct modifications to the background and perturbation dynamics, offering valuable test beds for cosmological analyses.

In the interacting framework, the total energy-momentum tensor is given by
\begin{equation}
T^{t}_{ij} = (\rho_{t} + p_{t})u_i u_j + p_{t}g_{ij}, \label{1}
\end{equation}
where $\rho_{t}$ and $p_{t}$ denote the total energy density and pressure, respectively. Explicitly, $\rho_{t} = \rho_{CDM} + \rho_\phi$, with CDM being pressureless ($p_{CDM}=0$), so that $p_{t} = p_\phi$. The corresponding total equation of state (EoS) is
\begin{equation}
p_{t} = w_{t}\rho_{t}, \label{2}
\end{equation}
where $w_{t}$ generally evolves with time.

The Einstein Field Equations in a flat FLRW background can then be written as
\begin{equation}\label{3}
\Big(\frac{\dot{a}}{a}\Big)^2 = \frac{1}{3}\,\rho_{t} 
= \frac{1}{3}\,(\rho_m + \rho_{\phi}) ,
\end{equation}
\begin{equation}\label{4}
\Big(\frac{\dot{a}}{a}\Big)^2 + 2\Big(\frac{\ddot{a}}{a}\Big) = -\,p_{t} = -\,p_{\phi},
\end{equation}
in units where $8\pi G=c=1$.

When dark energy is modeled by a scalar field $\phi$ with potential $V(\phi)$, it can be treated as a perfect fluid with canonical energy density and pressure \cite{Ratra1988,Caldwell1998,Tsujikawa2013}:
\[
\rho_{\phi} = \frac{1}{2}\dot{\phi}^{2} + V(\phi), 
\qquad 
p_{\phi} = \frac{1}{2}\dot{\phi}^{2} - V(\phi).
\]
These relations form the basis of quintessence models, in which the scalar field dynamics determine the dark energy equation of state. Explicitly,
\begin{equation}\label{5a}
\rho_{\phi}= \frac{\dot{\phi}^2}{2}+V(\phi),
\end{equation}
\begin{equation}\label{5b}
p_{\phi}= \frac{\dot{\phi}^2}{2}-V(\phi).
\end{equation}

In the interacting scenario, the conservation equations split into separate balance equations for baryons, dark matter, and dark energy. For non-relativistic baryons,
\begin{equation}\label{8}
\dot{\rho_b}+3\Big(\frac{\dot{a}}{a}\Big) \rho_b=0,
\end{equation}
with solution
\begin{equation}\label{6}
\rho_b = \frac{B}{a^{3}},
\end{equation}
where $B$ is an integration constant. For CDM, introducing a specific interaction term
\begin{equation}\label{7}
Q = 3\gamma H \rho_{dm}, 
\end{equation}
yields
\begin{equation}\label{8}
\rho_{dm} = \frac{D}{a^{3(\gamma - 1)}},
\end{equation}
with $D$ an integration constant. The DE density then satisfies
\begin{equation}
\dot{\rho}_\phi + 3\frac{\dot{a}}{a}\left(1 + \omega_\phi\right)\rho_\phi = -3H\gamma D\,a^{3(\gamma - 1)}, \label{9eos}
\end{equation}
where $\omega_\phi = p_\phi/\rho_\phi$. Equivalently, the evolution of the DE equation of state can be written as
\begin{equation}
\omega_\phi=-1-\frac{1}{3}a\frac{\rho_\phi'}{\rho_\phi}. \label{10}
\end{equation}

To close the system, we adopt a parametrization of the DE density as
\begin{equation}
\rho_\phi'+\beta f(a)\rho_\phi=0,\label{11}
\end{equation}
with
\[
f(a)=\beta^2\left[ 1+\frac{1}{a\sqrt{1+a^{2\beta}}\sinh^{-1}(1/a)^{\beta}} \right].
\]
Following our previous work \cite{ritikaEDSFD}, the choice of $f(a)$ is designed to induce a signature flip in the deceleration parameter $q$, ensuring a natural transition from an early decelerated phase (necessary for structure formation) to a late-time accelerated epoch consistent with current observations. Within the interacting framework, this parametrization allows us to investigate how energy exchange between DE and CDM influences the cosmic dynamics and addresses theoretical challenges such as the coincidence problem.

\textit{Parametrizing the dark energy (DE) density}, $\rho_{\phi}$, plays a crucial role in cosmological studies for several reasons. First, an explicit parametrization of $\rho_{\phi}$ allows one to directly quantify its impact on the expansion history of the Universe through the Friedmann equations. This enables precise evaluation of how variations in $\rho_{\phi}$ influence the Hubble parameter $H$ and the deceleration parameter $q$ \cite{huterer1999prospects, wang2004model}. Second, parametrized forms of $\rho_{\phi}$ provide a flexible and model-independent framework for fitting diverse observational datasets, leading to tighter constraints on dark energy properties and allowing potential time evolution to be probed \cite{maor2001dynamics, chevallier2001accelerating}. Third, treating $\rho_{\phi}$ as a dynamical quantity enables the modeling of complex dark energy behaviors, offering a viable approach to addressing fundamental issues such as the cosmic coincidence problem and accommodating a broad range of evolutionary scenarios \cite{linder2003exploring}. 

Furthermore, several theoretical models predict specific functional forms of $\rho_{\phi}$, and parametrizing this quantity allows these predictions to be confronted with observational data, thereby enabling existing theories to be tested, refined, or ruled out \cite{wang2004model}. Additionally, reconstructing the evolution of $\rho_{\phi}$ over cosmic time is essential for identifying key cosmological transitions, including the onset of dark energy domination after the matter-dominated era, and for characterizing the timing and nature of this transition \cite{maor2001dynamics, caldwell2003phantom}. Finally, a well-defined parametrization of $\rho_{\phi}$ facilitates the joint interpretation of multiple cosmological datasets within a unified framework, aiding in the resolution of current observational tensions—such as the discrepancy in measurements of the Hubble constant $H_0$—while promoting a consistent understanding of the Universe’s expansion history \cite{valentino2020hubble}.

The general solution of the above differential equation (\ref{11}) is given by
\begin{equation}
\rho_\phi=e^{-\beta a} sinh^{-1}\Big(\frac{1}{a}\Big)^{\beta},
\end{equation}\label{12}
where $\beta \in (0,1)$ is the model parameter.\\\\
Using the standard relation between the redshift \(z\) and the scale factor \(a\), we have
\[
\frac{a}{a_0} = \frac{1}{1+z}.
\]
Expressing the scalar field energy density \(\rho_\phi\) in terms of redshift \(z\),
\begin{equation}\label{13}
\rho_\phi(z)=e^{ \frac{-\beta}{1+z}} sinh^{-1}(1+z)^{\beta},
\end{equation}
and 
\begin{equation}\label{14}
\rho_{\phi_{0}}=e^{-\beta} sinh^{-1}(1).
\end{equation}    
Equations (\ref{13}) and (\ref{14}) yield
\begin{equation}\label{15}
\rho_\phi(z)= \frac{\rho_{\phi_{0}}}{sinh^{-1}(1)} e^{\frac{\beta z}{1+z}} sinh^{-1}(1+z)^{\beta},
\end{equation}
where $\rho_{\phi_{0}}$ is the present value of the DE energy density.
\begin{equation}
p_{\phi}=\left[ \left( 2q-1\right) H^{2}
\right] \text{.}  \label{30}
\end{equation}
where \[
q = -\frac{\ddot{a}a}{\dot{a}^{2}} = -1 - \frac{\dot{H}}{H^{2}},
\]
Also using equations (\ref{6}), (\ref{8}) and (\ref{15}) in equation (\ref{3}), we have
\begin{equation}\label{16}
 3H^2= B(1+z)^3+ D(1+z)^{3(1-\gamma)}+\frac{\rho_{\phi_{0}}}{sinh^{-1}(1)} e^{\frac{\beta z}{1+z}} sinh^{-1}(1+z)^{\beta},
\end{equation}
We define the density parameter \(\Omega=\frac{\rho}{\rho_c}\), where \(\rho_c=\frac{3H^2}{(8\pi G)^2}\) denotes the critical density. In normalized units, we set \(8\pi G=1\), simplifying the resulting expressions.\\\\
Equation (\ref{16}) in terms of density parameter of CDM and DE can be expressed as
\begin{equation}
\begin{aligned}
H(z) &= H_0 \bigg[ \Omega_{b_0}(1+z)^3 + \Omega_{dm_{0}}(1+z)^{3(1-\gamma)}\\
&\quad + \frac{\Omega_{\phi_{0}}}{\operatorname{arcsinh}(1)} \left( e^{\frac{\beta z}{1+z}} \operatorname{arcsinh}(1+z)^\beta \right) \bigg]^{1/2},
\end{aligned}
\label{17}
\end{equation}
where $\Omega_{b_0}=\frac{B}{3H_0^2}$, $\Omega_{dm_0}=\frac{D}{3H_0^2}$ and $\Omega_{\phi_0}=\frac{\rho_{\phi_{0}}}{3H_0^2}$ are the present values of baryonic matter, DM and DE density parameters respectively.

As a consistency check and to gain further insight into the interacting framework, we examine the conditions under which it reduces to the standard non-interacting model. In particular, if we set the interaction parameter $\gamma$  to zero (or let the coupling parameter approach zero), all energy transfer terms between the dark components vanish. Under this limit, the evolution equations recover their familiar non-interacting forms, and the model transitions smoothly to the standard scenario where CDM and DE evolve independently. This provides a useful benchmark, allowing us to compare and contrast the properties and predictions of the interacting model with those of the well-established non-interacting paradigm.
\begin{align*}\label{18}
H(z) &= H_0 \bigg[ \Omega_{b_0}(1+z)^3 + \Omega_{dm_{0}}(1+z)^3 \\
     &\quad + \frac{\Omega_{\phi_{0}}}{\operatorname{arcsinh}(1)} \left( e^{\frac{\beta z}{1+z}} \operatorname{arcsinh}(1+z)^\beta \right) \bigg]^{1/2},
\end{align*}
where \(\beta\) is the model parameter.\\\\
In the following section, we constrain the model parameters using a range of observational datasets. This approach identifies the best-fit current values of relevant cosmological parameters, enabling a more precise and comprehensive examination of their physical behavior.

\section{Methodology and Data Description}\label{sec_3}
In this section, we constrain the parameters of the Non-interacting and interacting dark energy models using the cosmological inference code \texttt{SimpleMC} \cite{simplemc,aubourg2015}, employing the Metropolis–Hastings Markov Chain Monte Carlo (MCMC) algorithm \cite{hastings1970monte} as the analyzer The convergence of the MCMC analysis was rigorously ensured by applying the Gelman–Rubin diagnostic $R - 1$ \cite{gelman1992inference} with a threshold of $R - 1 < 0.01$. In the post-visualization process, after obtaining the chain files of each parameter, we use the \texttt{GetDist} package \cite{lewis2025getdist} to visualize the parameter space of the model parameters.

For the statistical analysis, we calculate the Bayesian evidence $(\ln Z)$ using the \texttt{MCEvidence} package \cite{heavens2017marginal}, which provides a measure of how well a model fits the data. Model comparison is then performed using the Bayes factor $(B_{ab} = Z_a / Z_b)$, or equivalently, the difference in logarithmic evidence $(\Delta \ln Z)$. A higher value of $(\ln Z)$ corresponds to a more favored model. The strength of the evidence is assessed following the Jeffreys scale:
\begin{itemize}
    \item $(|\Delta \ln Z| < 1)$: Weak evidence
    \item $(1 \leq |\Delta \ln Z| < 3)$: Moderate evidence
    \item $(3 \leq |\Delta \ln Z| < 5)$: Strong evidence
    \item $(|\Delta \ln Z| \geq 5)$: Decisive evidence
\end{itemize}
During our analysis to obtain the parameter space of each model, we tested each model using different datasets such as Baryon Acoustic Oscillations, Unanchored Type Ia Supernovae, and the Compressed CMB Likelihood. Below, we provide more details about these datasets.
\begin{itemize}
    \item \textbf{Baryon Acoustic Oscillations:} First, we use the latest Baryon Acoustic Oscillation (BAO) measurements from the Dark Energy Spectroscopic Instrument (DESI) Data Release 2 \cite{karim2025desi}, which are extracted from different tracers: Bright Galaxy Sample (BGS), Luminous Red Galaxies (LRG1–3), Emission Line Galaxies (ELG1–2), Quasars (QSO), and Lyman-$\alpha$ forests. The BAO datasets are expressed in terms of the following dimensionless ratios: $D_M/r_d$, $D_H/r_d$, $D_V/r_d$, and $D_M/D_H$. Here, $D_H(z) = c/H(z)$ is the Hubble distance, $D_M(z) = c \int_0^z \frac{dz'}{H(z')}$ is the comoving angular diameter distance, and $D_V(z) \equiv [z D_M^2(z) D_H(z)]^{1/3}$ is the volume-averaged distance. The $r_d$ denotes the sound horizon at the drag epoch. In the flat $\Lambda$CDM model, $r_d = 147.09 \pm 0.20~\text{Mpc}$ \cite{aghanim2020planck}.
    
    \item \textbf{Unanchored Type Ia Supernovae}: Second, we use the Pantheon$^+$ sample \cite{brout2022pantheon}, which comprises 1701 light curves from 1550 Type Ia supernovae (SNe Ia). In our analysis, we exclude light curves with $z < 0.01$, as low-redshift data are subject to significant systematic uncertainties due to peculiar velocities. We marginalize over the $\mathcal{M}$ parameter; for further details, see Equations (A9--A12) of \cite{goliath2001supernovae}.
    
    \item \textbf{Compressed CMB Likelihood} Finally, we uses the two shift parameters, $R$ and $\ell_a$ \cite{wang2007observational}: $R \equiv \sqrt{\Omega_m H_0^2} \, D_M(z_*)$ , $\ell_a \equiv \pi \frac{D_M(z_*)}{r_s(z_*)} ,$ where $D_M(z_*)$ is the transverse comoving distance to the last-scattering surface and $r_s(z_*)$ is the comoving sound horizon at recombination. Typically, the CMB is compressed into a $3\times 3$ Gaussian likelihood for $\{R, \ell_a, \omega_b\}$. Since $\ell_a \propto \theta_s^{-1}(z_*)$, once $r_s(z_*)$ is calibrated (assuming $\Lambda$CDM  prior to recombination), $\ell_a$ determines $D_M(z_*)$, while $R$ constrains $\omega_m$. Together with $\omega_b$ and neglecting neutrinos, this effectively fixes $\omega_c$. This $3\times 3$ compression, known as the \texttt{Wang-Wang} likelihood in SimpleMC.
\end{itemize}
The posterior distributions of the cosmological parameters for both the non-interacting  and interacting dark energy models are obtained by combining the individual likelihoods $\mathcal{L}_{\text{tot}} = 
\mathcal{L}_{\text{BAO}} \times 
\mathcal{L}_{\text{SNe\,Ia}} \times 
\mathcal{L}_{\text{CMB}}.$ 

In our analysis, the radiation density parameter is computed using the relation 
$\Omega_{\text{rad},0} = 2.469 \times 10^{-5} \, h^{-2} 
\left( 1 + 0.2271\, N_{\text{eff}} \right)$ \cite{jarosik2011seven}, 
where $N_{\text{eff}} = 3.04$ is the standard effective number of relativistic species 
\cite{mangano2002precision}. The dark energy density is then determined from the flatness condition as $\Omega_{\phi} = 1 - \Omega_{m} - \Omega_{b} - \Omega_{\text{rad},0}.$ Hence, both $\Omega_{\text{rad},0}$ and $\Omega_{\phi}$ can be treated as redundant parameters 
in the parameter estimation process, rather than as independent free parameters. The choice of priors is mentioned in Table~\ref{tab_0}.

\begin{table}
\centering
\begin{tabular}{lll}
\hline
\textbf{Model} & \textbf{Parameter} & \textbf{Prior} \\
\hline
\multirow{3}{*}{\(\Lambda\)CDM} 
& \( \Omega_{m0} \) & \( \mathcal{U}[0.1, 0.5] \) \\
& \( \Omega_{b}h^2 \) & \( \mathcal{U}[0.02, 0.025] \) \\
& \( h \) & \( \mathcal{U}[0.4, 0.9] \) \\
\hline
\multirow{5}{*}{Interacting} 
& \( \Omega_{m0} \) & \( \mathcal{U}[0.1, 0.5] \) \\
& \( \Omega_{b}h^2 \) & \( \mathcal{U}[0.02, 0.025] \) \\
& \( \gamma \) & \( \mathcal{U}[-0.2, 0.2] \) \\
& \( \beta \) & \( \mathcal{U}[-0.5, 1.0] \) \\
& \( h \) & \( \mathcal{U}[0.4, 0.9] \) \\
\hline
\multirow{4}{*}{Non-Interacting} 
& \( \Omega_{m0} \) & \( \mathcal{U}[0.1, 0.5] \) \\
& \( \Omega_{b}h^2 \) & \( \mathcal{U}[0.02, 0.025] \) \\
& \( \beta \) & \( \mathcal{U}[-0.5, 1.0] \) \\
& \( h \) & \( \mathcal{U}[0.4, 0.9] \) \\
\hline
\end{tabular}
\caption{The table shows the priors used for each model in our analysis. The symbol $\mathcal{U}$ denotes a uniform prior distribution within the specified range.
use uniform priors, and $h \equiv H_0/100$.}\label{tab_0}
\end{table}

\begin{figure*}
\begin{subfigure}{.48\textwidth}
\includegraphics[width=\linewidth]{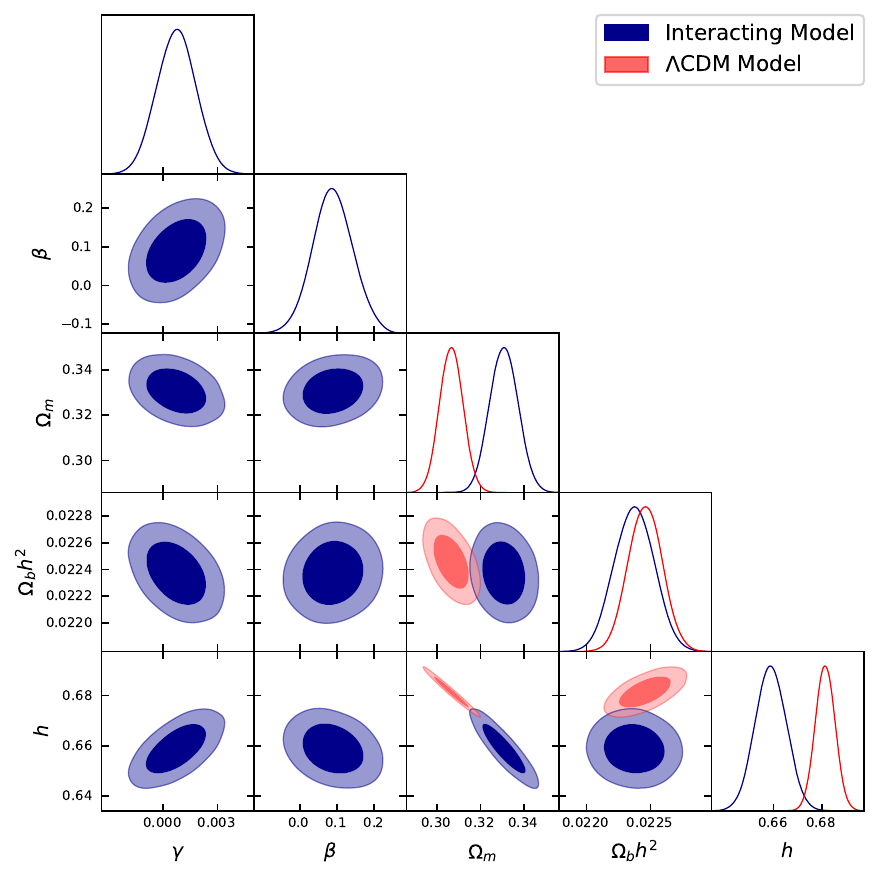}
    \label{fig_1a}
\end{subfigure}
\hfil
\begin{subfigure}{.48\textwidth}
\includegraphics[width=\linewidth]{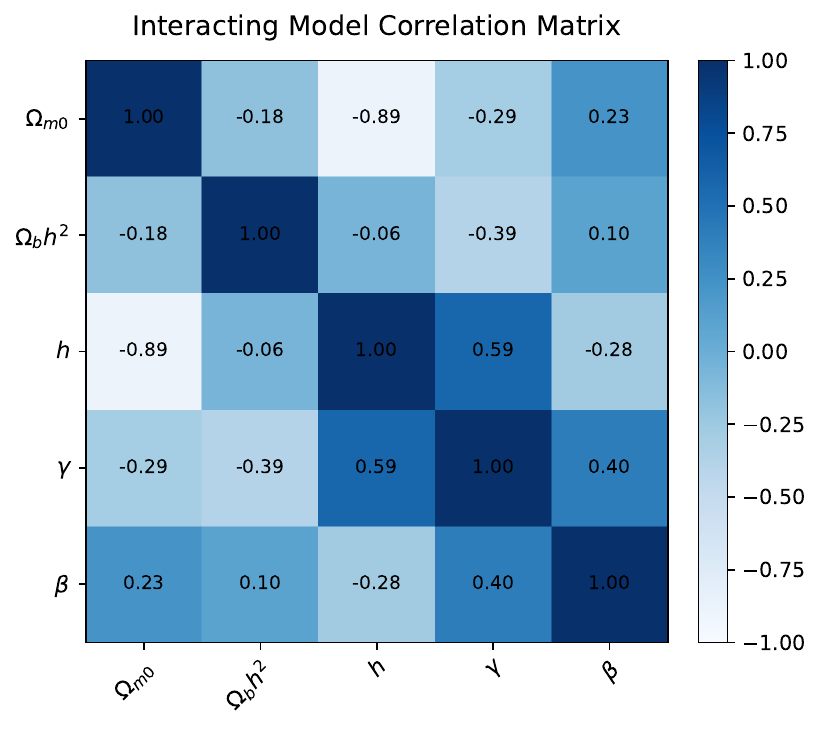}
    \label{fig_1b}
\end{subfigure}
\begin{subfigure}{.48\textwidth}
\includegraphics[width=\linewidth]{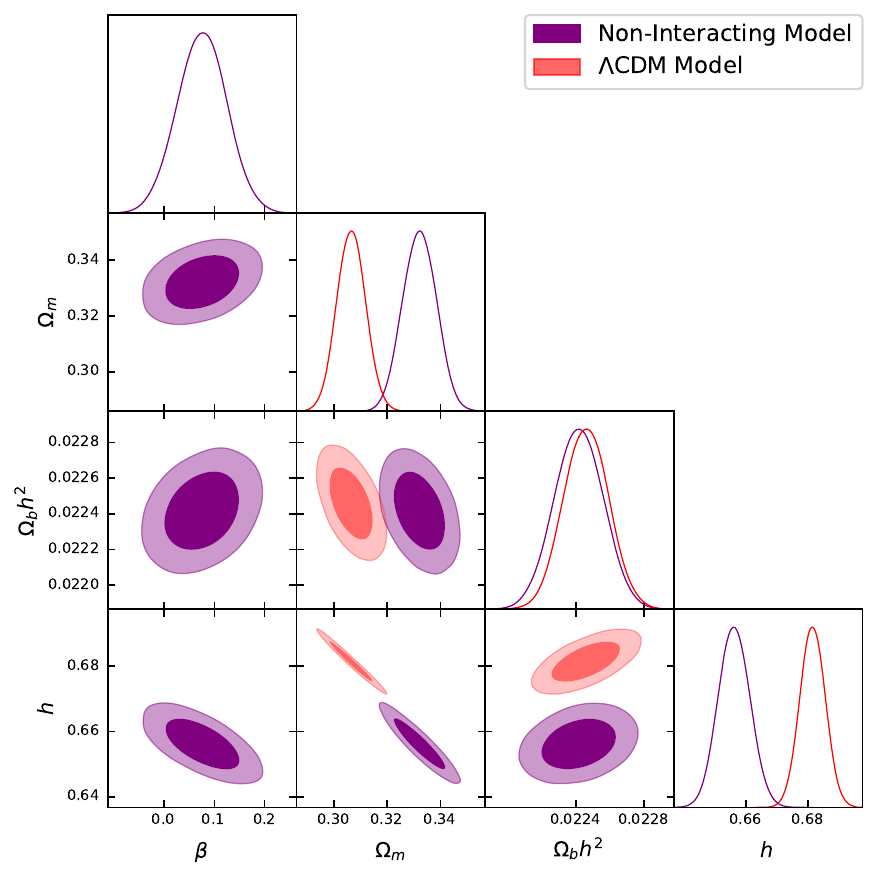}
    \label{fig_1a}
\end{subfigure}
\hfil
\begin{subfigure}{.48\textwidth}
\includegraphics[width=\linewidth]{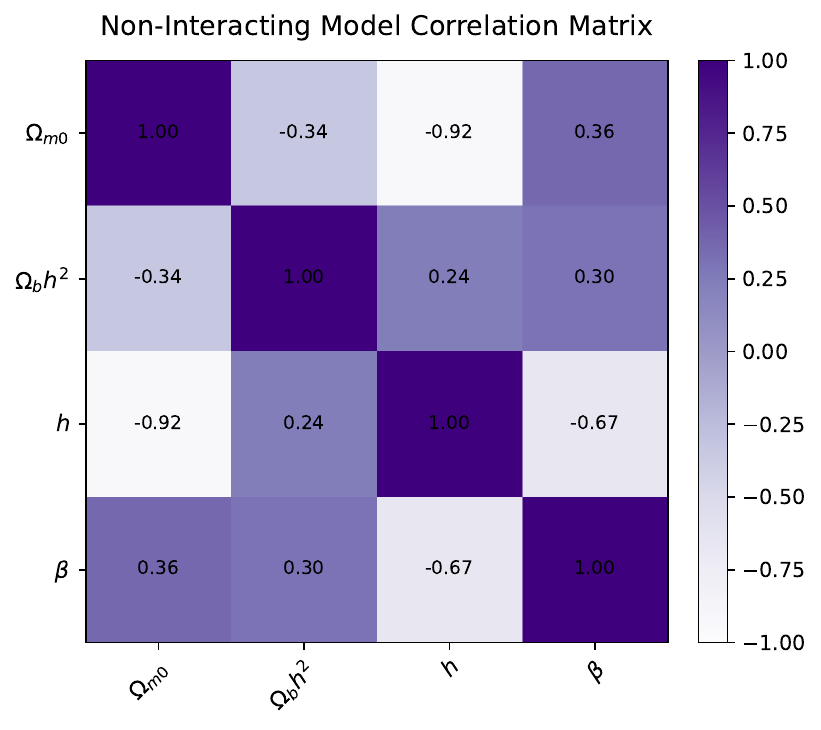}
    \label{fig_1b}
\end{subfigure}
\caption{The figure shows confidence contours at the 1\(\sigma\) and 2\(\sigma\) levels, based on constraints for the Non-Interacting, Interacting, and $\Lambda$CDM Models,  model}\label{fig_1}
\end{figure*}
\begin{table*}
\begin{tabular}{lcccccc}
\toprule
\textbf{Model} & $h$ & $\Omega_{m0}$ & $\Omega_{b}h^{2}$ & $\gamma$ & $\beta$  & $|\Delta \ln \mathcal{Z}|$ \\
\midrule
\textbf{$\Lambda$CDM} & $0.681 \pm 0.0041$ & $0.306 \pm 0.005$ & $0.02246 \pm 0.00013$ & -- & -- & 0 \\
\addlinespace[0.2cm]
\textbf{Interacting} & $0.659 \pm 0.0063$ & $0.330 \pm 0.006$ & $0.02237 \pm 0.00015$ & $0.0008 \pm 0.0010$ & $0.091 \pm 0.052$ & 5.00 \\
\addlinespace[0.2cm]
\textbf{Non-Interacting} & $0.656 \pm 0.0052$ & $0.332 \pm 0.006$ & $0.02241 \pm 0.00014$ & -- & $0.076 \pm 0.051$ & 0.72 \\
\bottomrule
\end{tabular}
\caption{Mean values, along with 68\% (1$\sigma$) credible intervals, for the standard $\Lambda$CDM model, Interacting, and Non-Interacting Models.}\label{tab_2}
\end{table*}

\subsection{MCMC and Statistical Results}
Fig.~\ref{fig_1}, shows the corner plots and correlation matrices for both the interacting and non-interacting  models. The upper panels correspond to the interacting model, while the lower panels show the results for the non-interacting model. The corner plots (left panels) shows the 2D marginalized confidence contours at the 68\% and 95\% confidence levels for the model parameters, together with their 1D posterior distributions along the diagonal. The contours shows the correlations between the cosmological parameters such as $\Omega_{m0}$, $\Omega_bh^2$, $h$, $\gamma$, and $\beta$, compared with the standard $\Lambda$CDM model (shown in red). The correlation matrices (right panels) quantify the degree of correlation between each pair of parameters. The color intensity indicates the strength and sign of the correlations, ranging from $-1$ (perfect anti-correlation) to $+1$ (perfect correlation). 

Table~\ref{tab_2} shows the mean values and 68\% (1$\sigma$) credible intervals of the cosmological parameters for the $\Lambda$CDM, Interacting, and Non-Interacting models. The $\Lambda$CDM model serves as the reference (baseline), providing standard cosmological values with $h = 0.681 \pm 0.0041$ and $\Omega_{m0} = 0.306 \pm 0.005$. The Hubble parameter in the Interacting model deviates from the $\Lambda$CDM value at about 2.93$\sigma$, while the Non-Interacting model differs by about 3.78$\sigma$. This indicates that both extended models predict a significantly lower $H_0$, thereby worsening the Hubble tension. The matter density parameter $\Omega_{m0}$ in the Interacting and Non-Interacting models deviates from the $\Lambda$CDM prediction by 3.07$\sigma$ and 3.33$\sigma$, respectively. Based on the Jeffreys scale, the Interacting model (\( |\Delta \ln \mathcal{Z}| = 5.00 \)) shows moderate evidence against the $\Lambda$CDM model, while the Non-Interacting model (\( |\Delta \ln \mathcal{Z}| = 0.72 \)) shows inconclusive evidence, remaining consistent with $\Lambda$CDM. In our analysis, we find $\gamma > 0$, indicating that dark matter decays into dark energy.

\section{Dynamical Systems Approach to Interacting Scalar Field Dark Energy Models}\label{sec_4}
We assume the matter content to consist of pressureless dust ($p_b = p_{dm} = 0$), and model dark energy with a canonical scalar field $\phi$ possessing an exponential potential $V(\phi)=\exp(\lambda\phi)$. In the interacting scenario, we introduce a coupling between the dark energy and dark matter via the interaction term $Q = 3\gamma H \rho_{dm}$, where $\rho_{dm}$ is the energy density of dark matter, $H$ is the Hubble parameter, and $\gamma$ is a dimensionless coupling parameter. This ansatz is commonly adopted in the literature for its mathematical tractability and physical interpretability as a deviation from the standard cold dark matter dilution \cite{das2018new,davari2018new}.

By comparing the phase portraits of interacting and noninteracting cases, we explore how even a mild interaction affects the nature and stability of critical points, thereby influencing the universe's evolution. This dynamical systems framework offers a powerful global perspective on all possible cosmological trajectories, not just those with specific initial conditions.

In the noninteracting case, we recover familiar critical points consistent with a matter-dominated epoch followed by a stable accelerated expansion mimicking a cosmological constant. When the interaction term $Q=3\gamma H\rho_{dm}$ is introduced, the nature and stability of critical points shift, potentially modifying the effective equation of state and altering the trajectory toward acceleration, depending on the sign and magnitude of $\gamma$.

These modifications carry important cosmological implications. Interacting dark energy models have been explored as possible resolutions to current observational tensions particularly the Hubble tension by introducing a coupling that subtly alters the background evolution and growth of structure \cite{das2018new,davari2018new}. Our phase-space analysis determines whether such models lead to a viable cosmic fate, including the existence of a stable, accelerated attractor, or instead result in pathological behaviors.

\subsection{Interacting scenario : $Q=3\gamma H \rho_{dm}$}
\begin{figure}[H]
		\centering
		\includegraphics[width=0.9\linewidth]{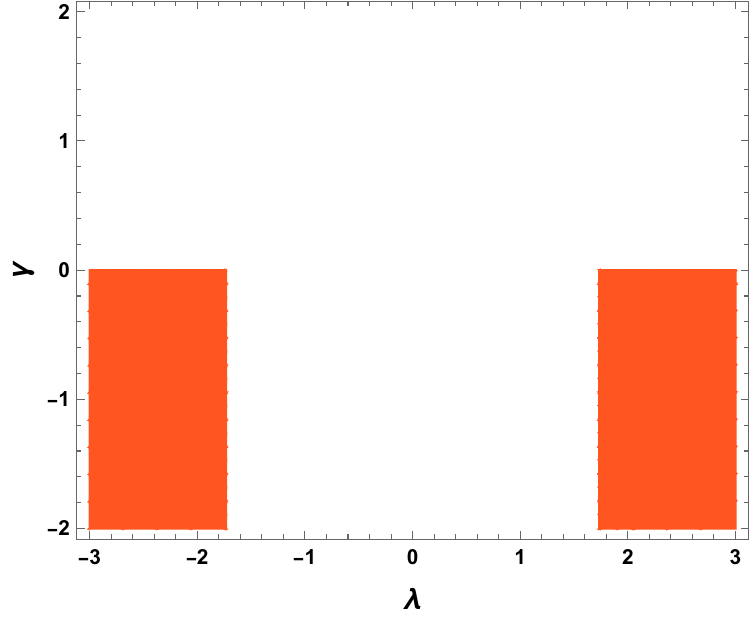}
		\caption{Region in $\left(\lambda,\gamma\right)$ space,where $A_{3\pm}$ exhibits stable behavior}
		\label{fig2 region plot}
	\end{figure}

\begin{figure}[H]
		\centering
		\includegraphics[width=0.85\linewidth]{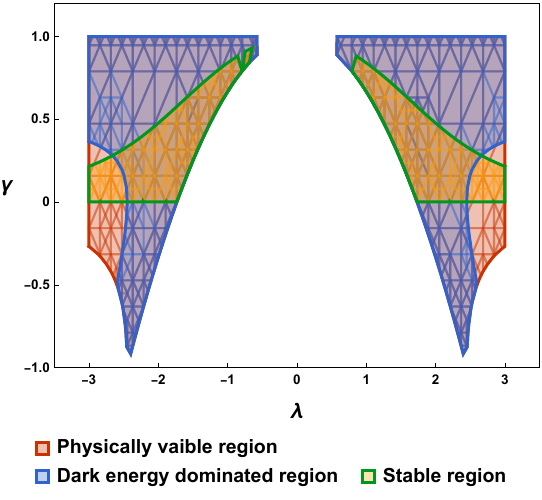}
		\caption{Region plot in $\lambda -\gamma$ parameter space corresponding to critical points $A_{5\pm}$}
		\label{fig3 region plot}
	\end{figure}

\begin{figure*}
\begin{subfigure}{.48\textwidth}
\includegraphics[width=\linewidth]{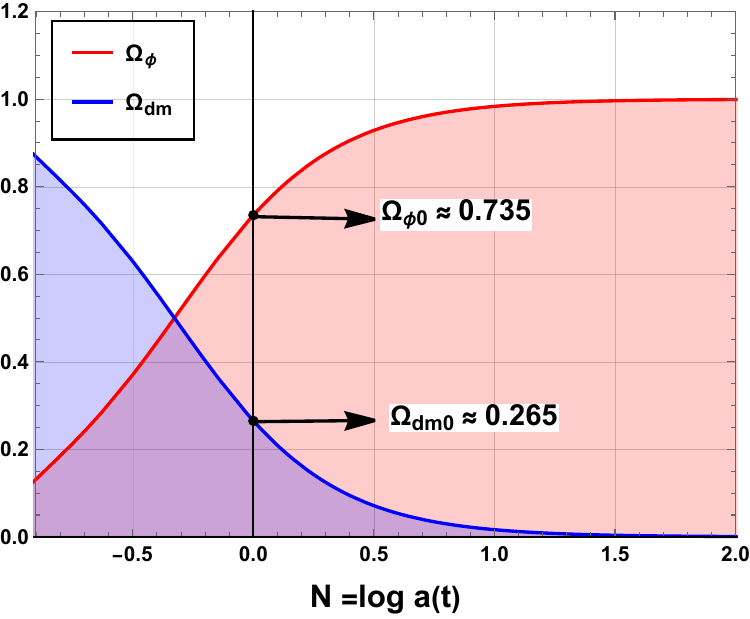}
    \label{4aa}
    \caption{Evolution of the background density parameters $\Omega_{dm}$ and $\Omega_\phi$}
\end{subfigure}
\hfil
\begin{subfigure}{.48\textwidth}
\includegraphics[width=\linewidth]{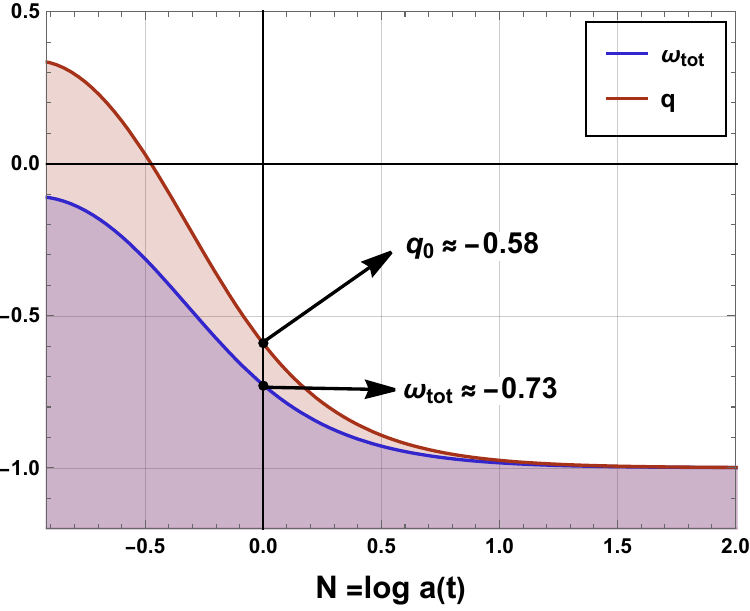}
    \label{4bb}
    \caption{Evolution of effective Eos parameter $\omega_{tot}$ and the deceleration parameter $q$}
\end{subfigure}
\caption{The numerical solutions of the equation \eqref{eq30}-\eqref{eq32}   describing the density parameters ($\Omega_{dm}$ and $\Omega_\phi$) and the Eos parameter ($\omega_{tot}$) together with the deceleration parameter (q) are presented.  The
vertical line $N = 0$ represents the current timeline, and $N > 0$ and $N < 0$ represent the future and past epochs,
respectively.}
\label{eos interaction}
\end{figure*}

We start our analysis for the interaction between dark energy and dark matter characterized by the interaction term $Q=3\gamma H\rho_{dm}$, where $\gamma$ is the coupling parameter that regulates the direction of energy exchange between dark sectors.For example, $\gamma<0$ indicates the energy exchange from dark energy to dark matter. On the contrary, $\gamma>0$ represents the energy transfer from dark matter to the dark energy component.This simple interaction model is constructed to achieve accelerated scaling attractors and it can also mitigate the coincidence problem.In order to construct the dynamical system and investigate the corresponding dynamics,we take the following  auxiliary variable
\begin{equation}\label{eq24}
    x=\frac{\dot{\phi}}{\sqrt{6}H},y=\frac{\sqrt{V\left(\phi\right)}}{\sqrt{3}H},z=\Omega_{dm}=\frac{\rho_{dm}}{3H^2},\Omega_b=\frac{\rho_b}{3H^2}
\end{equation}
One can see that the variable $x$ and $y$ are connected with the kinetic energy and the potential energy of the scalar field $\phi$, respectively, while $z$ and $\Omega_b$ are associated with dark matter and baryonic matter density, respectively.\\
Now ,using the expression of energy density and pressure component from equation \eqref{5a} and equation \eqref{5b} in the conservation equation \eqref{9eos}, we can write down the modified  Klein-Gordon equation for scalar field $\phi$ as
\begin{equation}
    \ddot{\phi}+3H\dot{\phi}+\frac{dV\left(\phi\right)}{d\phi}=-\frac{Q}{\dot{\phi}}
\end{equation}
Here "." represent the time derivative and $H=\frac{\dot{a\left(t\right)}}{a\left(t\right)}$ is usual Hubble parameter.\\
Under the transformation \eqref{eq24}, the first field equation \eqref{3} lead us to the following constrain
\begin{equation}
    x^2+y^2+z+\Omega_b=1
\end{equation}
In terms of the variable \eqref{eq24},the energy densities can be written as 
\begin{eqnarray}
 \Omega_{\phi}&=&\frac{\rho_\phi}{3H^2}=x^2 +y^2\\
 \Omega_{dm}&=&\frac{\rho_{dm}}{3H^2}=z\\
 \Omega_b&=&\frac{\rho_b}{3H^2}=1-z-x^2-y^2
\end{eqnarray}
The expression for scalar field Eos parameter is
\begin{equation}
    \omega_\phi=\frac{p_\phi}{\rho_\phi}=\frac{\frac{\dot{\phi}^2}{2}-V\left(\phi\right)}{\frac{\dot{\phi}^2}{2}+V\left(\phi\right)}=\frac{x^2-y^2}{x^2+y^2}
\end{equation}
Similarly, using the 2nd field equation \eqref{4} , we can write the expression for the total Eos parameter $\omega_{tot}$ and deceleration parameter $q$ as a function of dynamical variable as
\begin{eqnarray}
    \omega_{tot}&=&-1-\frac{2\dot{H}}{3H^2}=x^2-y^2\\
    q&=&\frac{1+3\omega_{tot}}{2}=\frac{1}{2}\left(3x^2-3y^2+1\right)
\end{eqnarray}
Finally, differentiating the variables \eqref{eq24} and using the conservation equation,we obtain the following autonomous dynamical system
\begin{eqnarray}
    \frac{d x}{d N}&=&-\frac{1}{2} 3x \left(y^2-x^2-1\right)-\frac{3 \gamma  z}{2 x}-3 x-\sqrt{\frac{3}{2}}\lambda y^2\label{eq30}\\
    \frac{d y}{d N}&=& \sqrt{\frac{3}{2}} \lambda  x y-\frac{1}{2}3y\left(y^2-x^2-1\right)\\
    \frac{d z}{d N}&=& 3 (\gamma -1) z-3 z \left(y^2-x^2-1\right)\label{eq32}
\end{eqnarray}
Here $N=\log a\left(t\right)$ is the logarithmic scale factor and $\lambda=\frac{V'\left(\phi\right)}{V\left(\phi\right)}$ is a constant parameter related to the scalar field potential.\\
In order to study the above dynamical system, the next step is to find out the critical points. The critical points can be obtained by solving the system of nonlinear equation $\frac{dx}{dN}=0$,$\frac{dy}{dN}=0$,$\frac{dz}{dN}=0$, from equation \eqref{eq30}-\eqref{eq32}. We associate each critical point with some particular epoch in the cosmic timeline by evaluating the value of density and eos parameters. Then, the stability properties of these points are investigated by calculating the eigenvalues of the Jacobian matrix, known as the linear stability theory. In Table \ref{table 1}, we have presented the critical points of the system \eqref{eq30}-\eqref{eq32}, with their variable existence condition. During the analysis, the critical points are considered to represent a real physical solution if it follows the standard existence condition, i.e. $0\leq \Omega_{dm},\Omega_b \leq 1$ and the expressions under the square root must be non-negative.
\begin{table}[h!]
 \centering
    \begin{tabular}{|c|c|c|c|c|}
        \hline
        &&&&\\
        Critical&$x$ & $y$ & $z$ & Existense \\
        Points&&&&condition\\
        \hline
        &&&&\\
        $A_{1\pm}$ & $\pm 1$ & $0$& $0$&Always\\
&&&&\\
\hline
&&&&\\
$A_{2{\pm}}$&$\pm \sqrt{- \gamma}$&$0$&$1+\gamma$&$\gamma<0$\\
&&&&\\
\hline
&&&&\\
$A_{3{\pm}}$&$-\frac{\sqrt{\frac{3}{2}}}{\lambda}$&$\pm\frac{\sqrt{\frac{3}{2}}}{\lambda}$&$0$&$\lambda\neq 0$\\
&&&&\\
\hline
&&&&\\
$A_{4{\pm}}$&$-\frac{\lambda}{\sqrt{6}}$ & $\pm \frac{\sqrt{6-\lambda^2}}{\sqrt{6}}$ &$0$&$\lambda^2<6$\\
&&&&\\
\hline
&&&&\\
$A_{5{\pm}}$&$\frac{\sqrt{\frac{3}{2}} (\gamma -1)}{\lambda }$ & $\pm \frac{\sqrt{\frac{3}{2} (\gamma -1)^2+\gamma  \lambda ^2}}{\lambda }$ &$\frac{(1-\gamma ) \left(3 \gamma +\lambda ^2-3\right)}{\lambda ^2}$& Figure \ref{fig3 region plot}\\
&&&&\\
\hline
\end{tabular}
\caption{Critical points along with their existence condition}
\label{table 1}
\end{table}
\\
\textbf{Critical point $A_{1\pm}$ :} The value of density and scalar field Eos parameters at these critical points are obtained as $\Omega_\phi=1$, $\Omega_{dm}=0$, $\Omega_b=0$ and $\omega_\phi=\omega_{tot}=1$, represents a completely dark energy dominated solution. For this particular class of solutions, the kinetic energy of the scalar field dominates over its potential energy. The value of the deceleration parameter is $q=\frac{1}{2}$, indicating the deceleration era. The set of eigenvalues of the Jacobian matrix corresponding to these critical points is $\{3,3\left(1+\gamma\right),\frac{1}{2}\left(6\pm\sqrt{6} \lambda\right)\}$.Due to the existence of a positive eigenvalue, these critical points cannot exhibit a stable cosmological solution. For $\gamma >-1\bigwedge \lambda >-\sqrt{6}$,$A_{1+}$ represents an unstable solution, and otherwise, it represents saddle behavior. Simillarly, for $\gamma >-1 \bigwedge \lambda <\sqrt{6}$,$A_{1-}$ represent a unstable behavior and otherwise it is a saddle point.\\\\
\textbf{Critical point $A_{2\pm}$ :} These two critical points exhibit a viable cosmological solution for $\gamma \leq 0$. The value of background cosmological density parameters at these critical points are $\Omega_\phi=-\gamma,\Omega_{dm}=1+\gamma$. Therefore, for $\gamma=-1$, these classes of solutions represent a completely dark energy-dominated solution.Also , the value of deceleration parameter is $q=\frac{1}{2}\left(1-3\gamma\right)$,exhibits an accelerated universe for $\gamma>\frac{1}{3}$.In terms of stability analysis,the eigenvalues of the Jacobian matrix are $\{-3\gamma,-3\left(1+\gamma\right),\frac{1}{2}\left(3-3\gamma+\sqrt{-6\gamma}\lambda\right)\}$.Under the constrain $\gamma<0$, dynamical stability cannot be achieved. But in the region,  $\gamma  (\gamma +1) \left(-\sqrt{6} \sqrt{-\gamma } \lambda +3 \gamma -3\right)>0$, these two critical can represent saddle behavior and otherwise these solutions are unstable node .\\\\
\textbf{Critical point $A_{3\pm}$ :}The value of density parameter corresponding to these critical points are $\Omega_\phi=\frac{3}{\lambda^2},\Omega_{dm}=1$.Therefore, for $\lambda=\pm \sqrt{3}$, these two critical points represent a completely dark energy-dominated universe.Also,the corresponding cosmological solution is associated with a decelerated matter-dominated epoch by a constant value of Eos parameter $\omega_\phi=\omega_{tot}=0$ and deceleration parameter $q=\frac{1}{2}$.The eigenvalues of the Jacobian matrix are $\{3\gamma,\frac{3 \left(-\lambda ^2-\sqrt{24 \lambda ^2-7 \lambda ^4}\right)}{4 \lambda ^2},\frac{3 \left(\sqrt{24 \lambda ^2-7 \lambda ^4}-\lambda ^2\right)}{4 \lambda ^2}\}$. Since the expressions of eigenvalues include $\gamma$ and $\lambda$, different combinations of these parameters affect the stability behavior. By numerical calculations of signatures of eigenvalue, we have presented a parametric region in figure \ref{fig2 region plot} , where $A_{3\pm}$ exhibits stable behavior and outside the shaded region, these critical points exhibit either saddle or unstable behavior.  
\noindent\textbf{Critical point $A_{4\pm}$ :}The cosmological solution corresponding to critical points $A_{4\pm}$ represent viable cosmological scenario only for $\lambda^2<6$.The value of background density parameters is $\Omega_\phi=1$,$\Omega_{dm}=0$, indicating complete domination of dark energy.The total Eos parameter and deceleration parameter are $\omega_{tot}=\omega_\phi=\frac{1}{3} \left(\lambda ^2-3\right)$ and $q=\frac{1}{2} \left(\lambda ^2-2\right)$ . Clearly, for constant potential, i.e., for $\lambda \to 0$ ,these critical points will represent the accelerated de-sitter era, and for $\lvert \lambda\rvert <\sqrt{2}$, it will represent the quintessence era.The eigenvalues of Jacobian matrix are $\left\{\frac{1}{2} \left(\lambda ^2-6\right),\lambda ^2-3,3 \gamma +\lambda ^2-3\right\}$.These two critical points are stable node only in the region $\left(\gamma \leq 0\land -\sqrt{3}<\lambda <\sqrt{3}\right)\lor \left(0<\gamma <1\land -\sqrt{3-3 \gamma }<\lambda <\sqrt{3-3 \gamma }\right)$, and otherwise they are either saddle or unstable node. \\\\
\textbf{Critical point $A_{5\pm}$ :}By considering the standard existence condition $0\leq\Omega_{dm}\leq 1$ and nonnegativity of expression under square root, we have found that these two critical points represent a valid cosmological scenario only in the region $\left(-3 \gamma ^2-2 \gamma  \lambda ^2+6 \gamma -3\right)\leq 0\land 0\leq -\frac{(\gamma -1) \left(3 \gamma +\lambda ^2-3\right)}{\lambda ^2}\leq 1$.At these points the values of background cosmological parameters are $\Omega_{\phi}=\frac{3 \gamma ^2+\gamma  \left(\lambda ^2-6\right)+3}{\lambda ^2}$,$\Omega_{dm}=-\frac{(\gamma -1) \left(3 \gamma +\lambda ^2-3\right)}{\lambda ^2}$, $\omega_{tot}=-\gamma $,$q=\frac{1}{2} (1-3 \gamma )$.One can note that for different combinations of $\left(\lambda,\gamma\right)$, these solutions will replicate various cosmological eras. For example,$\gamma=1$ can lead us to the de-sitter solution and if $\frac{1}{3}<\gamma <1$, the solution represents an accelerated quintessence era, and the phantom epoch can be obtained for $\gamma >1$.In Figure\ref{fig3 region plot}, dark energy dominated region corresponding to $A_{5\pm}$ is presented. We do not mention the eigenvalues for $A_{5\pm}$ in the manuscript, since they are complicated expressions of $\lambda$,$\gamma$. But we have numerically analyzed the eigenvalues in order to find the stability and in the Figure \ref{fig3 region plot} a stable region in $\lambda - \gamma$ parameter space, corresponding to $A_{5\pm}$ is presented.
    The phase space diagram for critical points $A_{i\pm}$,$\left(i=1,2,..,5\right)$ are presented in Figure \ref{fig4a},\ref{fig4b},\ref{fig4c} respectively.For stable critical points, the trajectories near the critical points converge toward the points, while for unstable critical points, the neighborhood trajectories are moving away from the points, and for saddle-like critical points, some trajectories are attracted, and others are repelled from them.\\\\
   In order to study the dynamical framework of the considered cosmological model, we examine the evolution of key background parameters—namely, the matter density $\Omega_{dm}$, dark energy density $\Omega_\phi$, effective equation of state parameter $\omega_{tot}$, and deceleration parameter $q$ respectively, in Fig.\ref{eos interaction}. Some recent astronomical observations reveals that our Universe is nearly flat, with the current values of background density parameters being $\Omega_{\phi}\approx0.7$ and $\Omega_{dm}\approx0.3$. We have numerically integrated the dynamical system \eqref{eq30}-\eqref{eq32} with suitable initial condition to trace the cosmic evolution across different epochs, as shown in the left panel of Fig.\ref{eos interaction}. The vertical line at $N=0$ marks the current epoch, while $N<0$ and $N>0$ denote past and future epochs, respectively. As it can be seen from Fig.\ref{eos interaction} that the present values $\Omega_{dm0}\approx0.265$ and $\Omega_{\phi0}\approx0.735$ align well with current observations. Initially, the matter density dominated over dark energy; however, as the Universe evolved, $\Omega_{dm}$ decreased while $\Omega_\phi$ increased steadily. In the asymptotic future, $\Omega_\phi\to1$ and $\Omega_{dm}\to0$, indicating a complete dark energy–dominated phase. Similarly, it is evident from the right panel of Fig.\ref{eos interaction}, the current values of the Eos parameter is obtained as $\omega_{tot}\approx-0.73$, representing a quintessence-like epoch and the deceleration parameter $q$ shows a transition from early deceleration ($q>0$) to late-time acceleration ($q<0$). The present value $q_0=-0.58$ , which satisfies the observational data, confirms the Universe’s ongoing accelerated expansion.
\subsection{Non-interaction scenario : $Q=0$}
In this subsection, a dynamical system analysis has been performed for the scalar field dark energy model, where the dark sectors do not interact with each other.To construct the corresponding dynamical system and its further analysis,we can consider the standard non-interaction model as a particular case of interaction model by setting the interaction parameter $\gamma=0$.The dynamical variables \eqref{eq24} that have been used to analyze the previous interaction scenario, are applicable to this non-interaction model also.\\
The dynamical system corresponding to the non-interaction scenario is given below
\begin{eqnarray}
    \frac{d x}{d N}&=&-\frac{1}{2} 3 x \left(y^2-x^2-1\right)-3 x-\sqrt{\frac{3}{2}} \lambda  y^2\label{eq43}\\
    \frac{d y}{d N}&=&\sqrt{\frac{3}{2}} \lambda  x y-\frac{1}{2} 3 y \left(y^2-x^2-1\right)\label{eq44}\\
    \frac{d z}{d N}&=&-3 z \left(y^2-x^2-1\right)-3z\label{eq45}
\end{eqnarray}
\begin{table}[h!]
 \centering
    \begin{tabular}{|c|c|c|c|c|}
        \hline
        &&&&\\
        Critical point&$x$ & $y$ & $z$ & Existence condition\\
        &&&&\\
        \hline
        &&&&\\
        $B_{1}$&$0$&$0$&Any&Always\\
&&&&\\
\hline
&&&&\\
$B_{2{\pm}}$&$\pm 1$&0&Any&Always\\
&&&&\\
\hline
&&&&\\
$B_{3{\pm}}$&$-\frac{\sqrt{\frac{3}{2}}}{\lambda}$&$\pm \frac{\sqrt{\frac{3}{2}}}{\lambda}$&Any&$\lambda\neq 0$\\
&&&&\\
\hline
&&&&\\
$B_{4{\pm}}$&$-\frac{\lambda}{\sqrt{6}}$ & $\pm \frac{\sqrt{6-\lambda^2}}{\sqrt{6}}$ &$0$&$\lambda^2<6$\\
&&&&\\
\hline
\end{tabular}
\caption{Critical points along with their existence condition}
\label{table 2}
\end{table}
\\
\textbf{Critical point $B_{1}$ :} At this solution, the kinetic and potential energy of the scalar field vanishes with the dark energy density $\Omega_{\phi}=0$ , indicates that the scalar field does not contribute to the cosmological dynamics. Also the value of total Eos parameter $\Omega_{tot}=0$ and deceleration parameter $q=\frac{1}{2}$ , represents a decelerated matter-dominated era.The eigenvalues of the Jacobian matrix are $\left\{-\frac{3}{2},\frac{3}{2},0\right\}$.Due to the presence of both positive and negative eigenvalues, this solution cannot be stable and always represents the saddle behavior. \\
\textbf{Critical point $B_{2\pm}$ :} In these classes of critical points, the potential energy of the scalar field is absent and the kinetic energy drives the complete dynamics and leads to a stiff matter solution. Although a stiff-matter EoS $\omega_{tot}=\omega_{\phi}=1$ is not viable at the classical macroscopic level, these solutions should only be important in the early Universe. The value of the deceleration parameter $q=2$ , represents a decelerated Universe.The Jacobian matrix has the eigenvalues $\left\{3,3,\frac{1}{2} \left(6\pm\sqrt{6} \lambda \right)\right\}$. \\\\

\textbf{Critical point $B_{3\pm}$ :} In these solution, the kinetic and potential energy of the scalar field equally contribute to the evolution dynamics.The dark energy density parameter $\Omega_\phi=\frac{3}{\lambda^2}$,thus for $\lambda=\pm\sqrt{3}$ dark energy dominated solution can be obtained .Also, the value of scalar field Eos and total Eos parameter coincide with constant value $\omega_\phi=\omega_{tot}=0$. Hence, these classes of solutions identify the cosmological evolution as a combined effect of both the matter and the scalar field but behave as a completely matter dominated era. The eigenvalues of the Jacobian matrix are $\left\{0,\frac{3 \left(-\lambda ^2-\sqrt{24 \lambda ^2-7 \lambda ^4}\right)}{4 \lambda ^2},\frac{3 \left(\sqrt{24 \lambda ^2-7 \lambda ^4}-\lambda ^2\right)}{4 \lambda ^2}\right\}$.For $\lambda^2 >3$, these solutions can represent a stable solution; otherwise, they are either unstable or saddle. \\
\textbf{Critical point $B_{4\pm}$ :}These classes of critical points were also present in the interaction model as $A_{4\pm}$. These solutions always represent a dark energy-dominated Universe as $\Omega_\phi=1$ and $\Omega_{dm}=0$.The Eos parameter values are $\omega_{tot}=\omega_\phi=\frac{\lambda^2-3}{3}$.Therefore for $\lambda^2 <2$, they represent accelerating solutions, and in the limit $\lambda \to 0$, the solution reduces to the de-sitter Universe where the dark energy acts as a cosmological constant.The eigenvalues of Jacobian matrix are $\left\{\frac{1}{2} \left(\lambda ^2-6\right),\lambda ^2-3,\lambda ^2-3\right\}$.For a dynamically stable solution we must have $-\sqrt{3}<\lambda <\sqrt{3}$ and for saddle like solution $-\sqrt{6}<\lambda <-\sqrt{3}\lor \sqrt{3}<\lambda <\sqrt{6}$ .\\\\

\begin{figure*}
\begin{subfigure}{.48\textwidth}
\includegraphics[width=\linewidth]{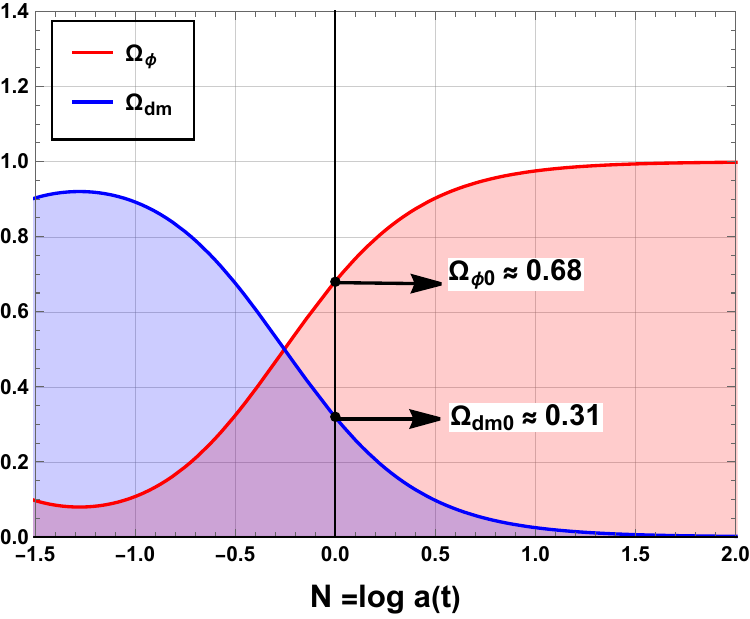}
    \label{fig5aa}
\end{subfigure}
\hfil
\begin{subfigure}{.48\textwidth}
\includegraphics[width=\linewidth]{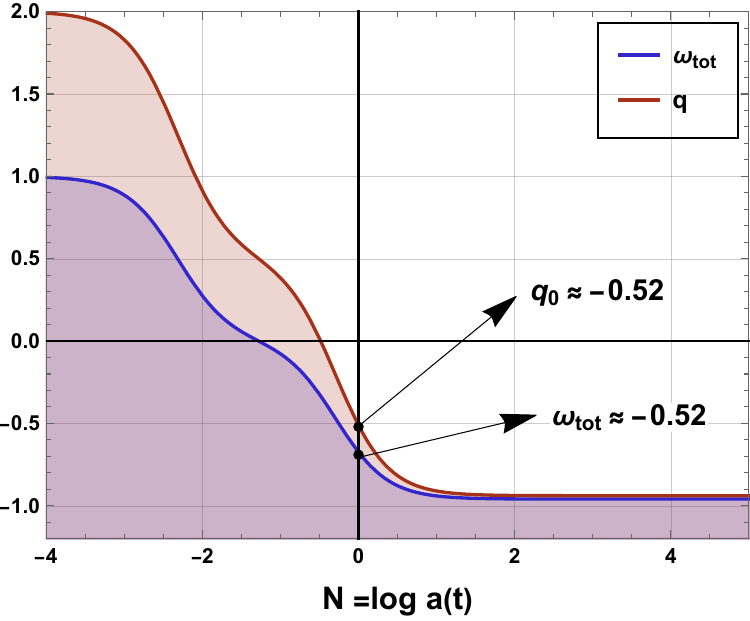}
    \label{fig5bb}
\end{subfigure}
\caption{The numerical solutions of \eqref{eq43}, \eqref{eq44} and \eqref{eq45}  describing the density parameters and the Eos parameter ($\omega_{eff}$) together with the deceleration parameter (q) are presented.  The
vertical line N = 0 represents the current timeline, and N $>$ 0 and N $<$ 0 represent the future and past epochs,
respectively.}\label{eos noninteraction}
\end{figure*}

\begin{figure*}[htb]
\begin{subfigure}{.32\textwidth}
\includegraphics[width=\linewidth]{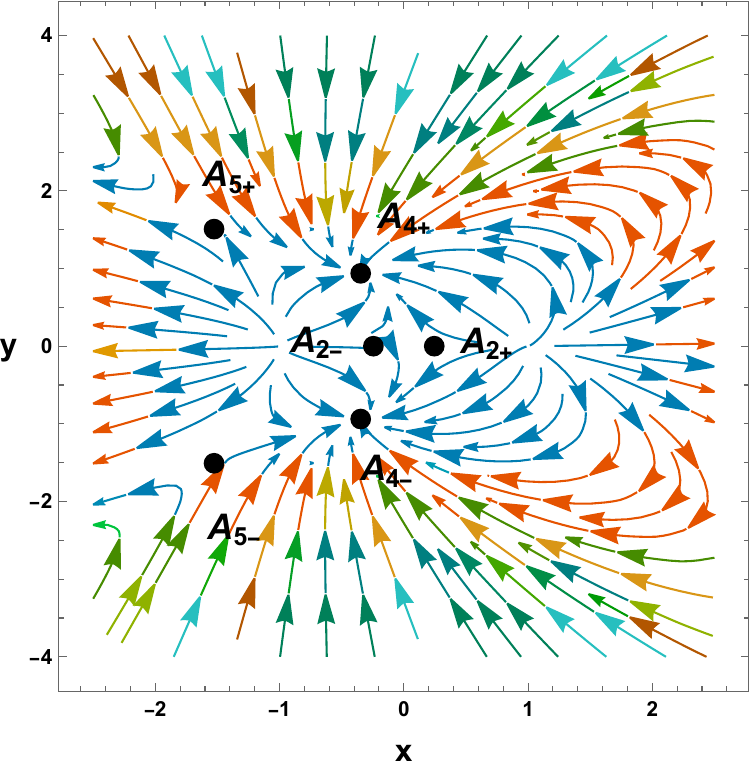}
    \caption{Projection of phase space on $x-y$ plane}
    \label{fig4a}
\end{subfigure}
\hfil
\begin{subfigure}{.32\textwidth}
\includegraphics[width=\linewidth]{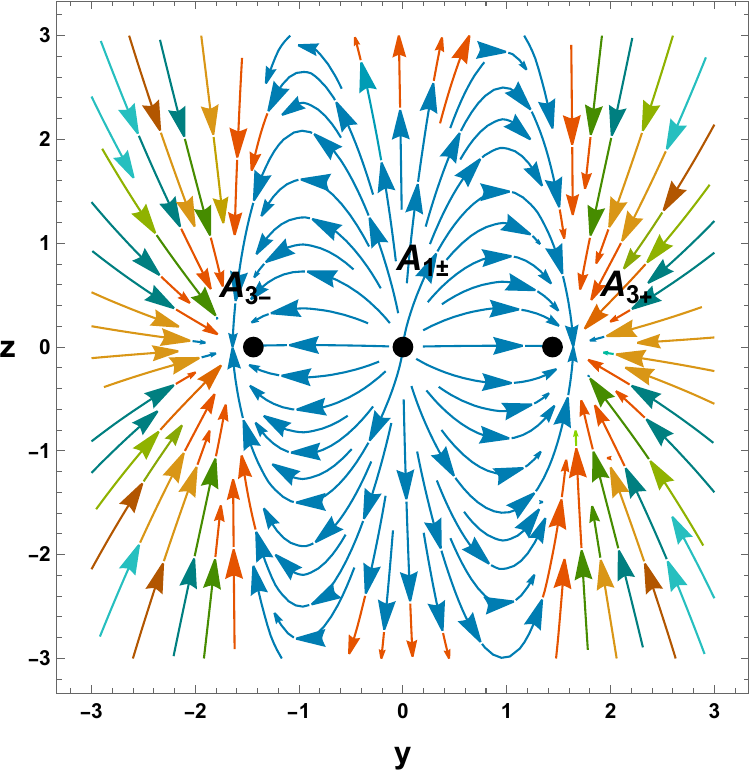}

    \caption{Projection of phase space on $y-z$ plane}
    \label{fig4b}
\end{subfigure}
\hfil
\begin{subfigure}{.32\textwidth}
\includegraphics[width=\linewidth]{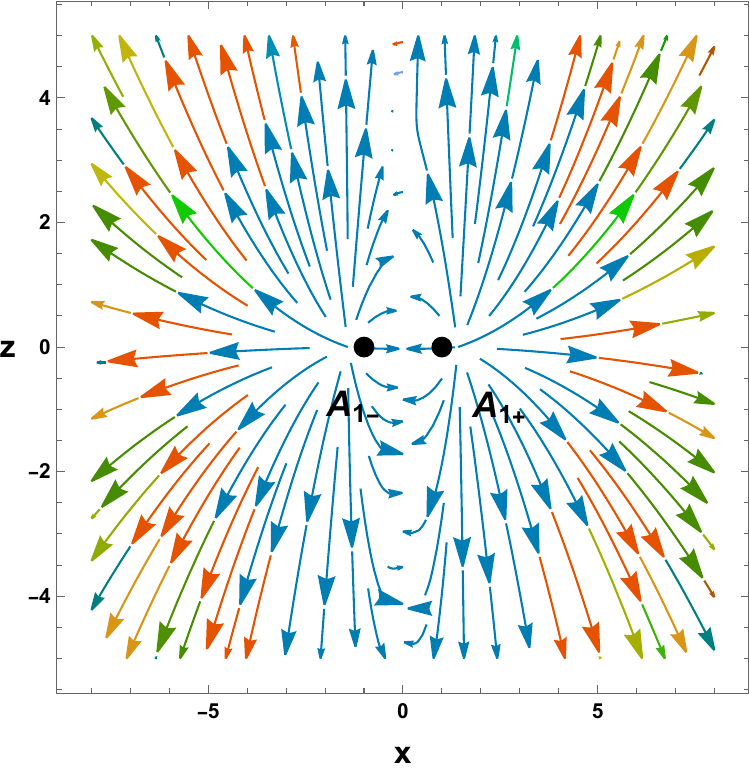}
    \caption{Projection of phase space on $x-z$ plane}
    \label{fig4c}
\end{subfigure}
\hfil
\begin{subfigure}{.33\textwidth}
\vspace{1cm}
\includegraphics[width=\linewidth]{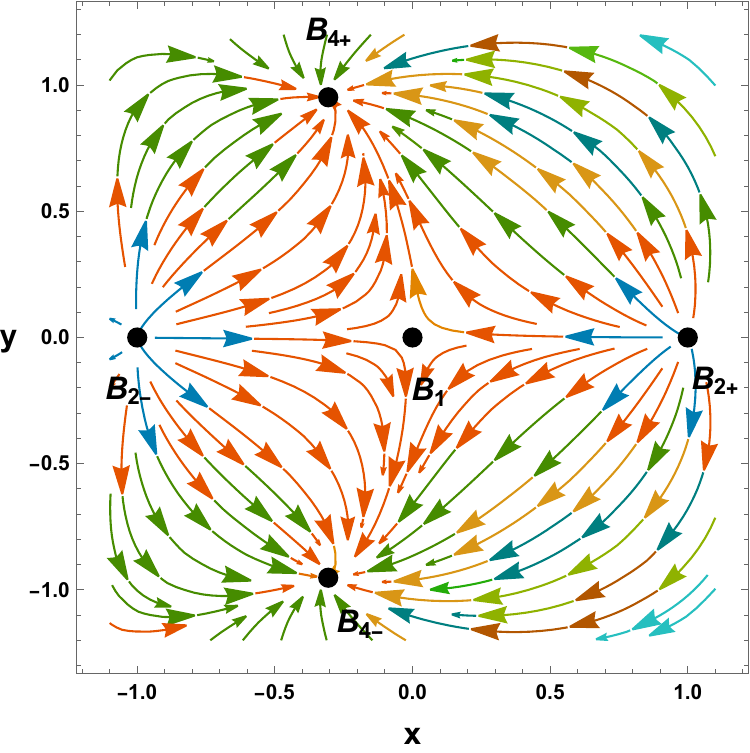}
    \caption{Projection of phase space on $x-y$ plane}
    \label{fig4d}
\end{subfigure}
\hfil
\begin{subfigure}{.32\textwidth}
\includegraphics[width=\linewidth]{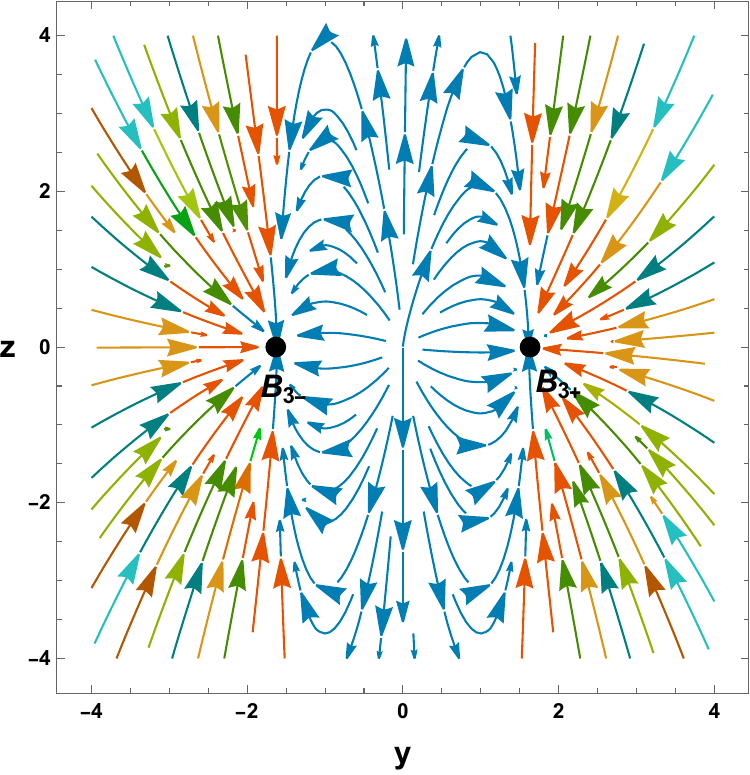}
    \caption{Projection of phase space on $y-z$ plane}
    \label{fig4e}
\end{subfigure}
\begin{subfigure}{.32\textwidth}
\includegraphics[width=\linewidth]{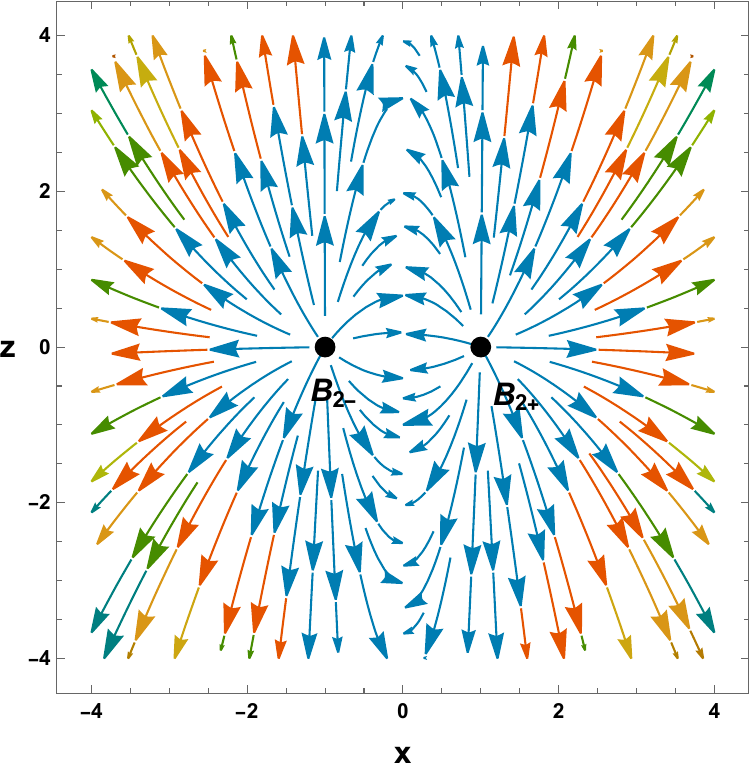}
    \caption{Projection of phase space on $x-z$ plane}
    \label{fig4f}
\end{subfigure}
\caption{Projection of phase space diagram for interaction model (upper panel) and non-interaction model (lower panel)}\label{}
\end{figure*}

The phase space diagram corresponding to the critical points $B_{i\pm},\left(i=1,2,..,4\right)$ have presented in Figure \ref{fig4d},\ref{fig4e},\ref{fig4f} respectively.In both the interaction and non-interaction scenarios, the critical points represent some particular epoch in the cosmic timeline. In the view of dynamical system analysis, an ideal cosmological model should represent the following cosmological era Inflation $\to$ matter/radiation era $\to$  late acceleration era, for some of the critical points. For both the interaction and non-interaction model of scalar field dark energy, the expression of critical points contains the terms $\lambda$ and $\gamma$ representing the effect of scalar field potential and interaction in the cosmological dynamics. We have also noticed that by considering various combinations of $\lambda$ and $\gamma$ , the critical points can successfully exhibit different cosmological epochs starting from matter domination to the late time acceleration era.In both scenarios, the late time acceleration behavior is similar to the $\Lambda$CDM model.\\\\
The evolution of density parameters $\Omega_{dm}$ and $\Omega_\phi$ along with the Eos parameter $\omega_{tot}$ and the deceleration parameter $q$ is presented in Fig.\ref{eos noninteraction}. From, the left panel of Fig.\ref{eos noninteraction}, it is evident that the current values of density parameters corresponding to the non-interaction scenario is obtained as $\Omega_{dm}\approx0.29$ and $\Omega_{\phi}\approx0.69$ respectively, which indicate the current day dark energy domination over dark matter sector. Similarly, from the right panel of Fig.\ref{eos noninteraction}, the value of the Eos parameter corresponding to the non-interaction model is obtained at $\omega_{tot}\approx-0.71$, which indicates a quintessence-type dark energy scenario. Additionally, the value of the deceleration parameter is obtained as $q\approx-0.54$. This negative value of the deceleration parameters indicates the current accelerated expansion, and the numerically obtained values are aligned with observational data.


\section{Conclusion}\label{sec_5}
In this work, we presented a detailed dynamical systems and observational analysis of an interacting scalar field dark energy (SSFDE) models with an exponential potential and dark-sector coupling \(Q=3\gamma H\rho_{dm}\). The cosmological evolution was studied in both interacting and non-interacting scenarios, and the resulting autonomous system revealed multiple critical points corresponding to radiation-like, stiff-fluid, matter-dominated, and late-time accelerated epochs. The stability of these points depends sensitively on the coupling parameter \(\gamma\) and the potential slope \(\lambda\), yielding well-defined regions of physical viability in the \((\lambda,\gamma)\) parameter plane.

Among the critical points, \(A_{4\pm}\) and \(A_{5\pm}\) exhibit stable accelerated attractor behavior within constrained domains, ensuring a graceful transition from matter domination to late-time acceleration. Numerical phase-space evolution provides consistency with current cosmological observations, accurately reproducing present-day values: \(\Omega_{\phi 0}\approx0.735\), \(\Omega_{dm0}\approx0.265\), total equation of state \(\omega_{\mathrm{tot}}\approx -0.73\), and deceleration parameter \(q_0\approx -0.58\), demonstrating that the model naturally yields a realistic expansion history without fine-tuning of initial conditions.

The cosmological parameters were constrained using the latest observational datasets, including DESI DR2 BAO measurements, Pantheon+ Type Ia supernovae, and the compressed CMB likelihood. A Metropolis Hastings MCMC analysis with Bayesian evidence evaluation showed that the interacting dark energy model provides a competitive fit to data, and in comparison with \(\Lambda\)CDM, shows improved capability in alleviating the \(H_0\) and sound horizon (\(r_d\)) tensions. The Jeffreys scale assessment validates that the interaction parameter \(\gamma\) is both statistically viable and physically meaningful, governing the direction of energy transfer (\(\gamma>0\): dark matter \(\rightarrow\) dark energy, \(\gamma<0\): dark energy \(\rightarrow\) dark matter), and thereby offering an interpretable mechanism for addressing the cosmic coincidence problem.

The dynamical analysis further reveals that while the non-interacting case admits standard matter-to-acceleration transition, interaction introduces new stability structures and expands the physically admissible parameter space, enabling late-time attractors that are otherwise absent or unstable. No ghost or early-time instabilities were encountered within the viable domain, ensuring physical consistency of the model.

In summary, this study establishes that SSFDE with dark-sector coupling admits stable accelerated attractors and the predicted cosmological parameters align with current measurements. Moreover, dynamical behavior naturally resolves late-time acceleration without external modification and the model provides easing of existing cosmological tensions in \(H_0\) and \(r_d\). These results support interacting scalar field dark energy as a robust, observationally consistent, and dynamically stable extension of \(\Lambda\)CDM, offering a well-motivated theoretical framework.

\bibliographystyle{elsarticle-num}
\bibliography{mybib.bib,desi}

\end{document}